\documentclass[preprint]{aastex}

\usepackage{graphicx}
\usepackage{epsf}
\usepackage{natbib}

\newcommand{\simlt}{\lower.5ex\hbox{$\; \buildrel < \over \sim \;$}}

\providecommand{\sorthelp}[1]{}

\begin{document}

\title{Variations between Dust and Gas in the Diffuse Interstellar Medium. 2. Search for Cold Gas}

\shorttitle{Dark, cold atomic gas}

\author{William T. Reach}
\affil{Universities Space Research Association, MS 232-11, Moffett Field, CA 94035, USA}
\email{wreach@sofia.usra.edu}
\and
\author{Carl Heiles}
\affil{Astronomy Department, University of California, Berkeley, CA 94720, USA}
\and
\author{Jean-Philippe Bernard}
\affil{Universit\'e de Toulouse, Institut de Recherche en Astrophysique et Plan\'etologie, F-31028 Toulouse cedex 4, France}

\begin{abstract}
The content of interstellar clouds, in particular the inventory of diffuse molecular gas, remains uncertain. 
We identified a sample of isolated clouds, approximately 100 solar masses in size, and
used the dust content to estimate the total amount of gas. In Paper 1, the total inferred gas content
was found significantly larger than that seen in 21-cm emission measurements of H~I. In this paper we 
test the hypothesis that the apparent excess `dark' gas is cold H~I, which would be evident in absorption
but not in emission due to line saturation. The results show there is not enough 21-cm absorption toward
the clouds to explain the total amount of `dark' gas. 
\end{abstract}


\section{Introduction}

Significant amounts of `dark' gas, whose amount is not accurately measured by the widely-used interstellar tracers (the 21-cm line of atomic hydrogen or 2.7-mm line of CO), have been inferred from $\gamma$-ray \citep{grenier05}, 
far-infrared \citep{hrk88,reach98}, and extinction \citep{paradis12} observations of the diffuse interstellar medium and the Magellanic Clouds \citep{bernard08,romanduval10}.
A survey of [C~II] emission provided another, independent tracer confirming the presence of CO-dark
gas \citep{langer14}.
The possible presence of `dark gas' is important for understanding the inventory of 
baryonic material within galaxies---in particular, the amount of molecular gas from
which stars form. Using dust as a tracer of the total amount of interstellar material,
it was found that the total amount of interstellar material in galaxies 
decreases with time as galaxies evolve \citep{scoville14}. 
Molecular gas at  $> 30$ K is predicted to be `CO-dark' \citep{glover16}, so the typical interstellar medium probed by UV absorption lines, with temperatures averaging $77$ K \citep{savage77,sofia05} belongs to the `dark gas' phase.
Directly measuring the amount of warm ($>100$ K) H$_2$ and extrapolating to the total amount of H$_2$ in galaxies suggests molecular masses could even be 100 times larger, in some galaxies, than would be derived from CO observations alone
\citep{togi16}.

In Paper 1 \citep{reach15}, we studied a set of clouds in the local interstellar medium
and assessed evidence for the presence of `dark gas' within them, by
comparing the amount of dust seen with {\it Planck} (far-infrared and sub millimeter
continuum) to the amount of gas seen with Arecibo (21-cm line emission). 
We found that these 
$\sim 100 M_\odot$ clouds frequently contain factors of $\sim 2$--3 times more dust 
per unit gas than is present in the diffuse ISM. 
Inventories of abundance of the elements
already use up all the heavy elements in dust in the diffuse interstellar medium
\citep{jenkins09}, so the
excess dust seen in the clouds must be either due to unseen `dark' gas or to changes
in dust properties that cause us to overestimate the amount of dust. 

In this paper, we explore the hypothesis that the unseen gas in interstellar clouds is 
cold, atomic gas. Such gas would have optically-thick 21-cm line emission, which would cause
us to underestimate its presence from 21-cm line emission measurements. 
A recent paper has found it plausible that most of the purported `dark gas' in 
the interstellar medium is actually atomic gas that is optically thick in 
the 21-cm line \citep{fukui15}. Now that we have a well-characterized set of 
nearby interstellar clouds with quantified amounts of supposed `dark gas,' we 
have followed up with a directed study of the 21-cm optical depth to determine whether
the hypothesis applies.

\section{Observations \label{obssec}}

\subsection{Target selection}
The interstellar clouds for this study are those used in Paper 1; they are ~degree-sized clouds readily evident in the {\it Planck} far-infared and Arecibo/GALFA 21-cm surveys as being relatively isolated structures separated from the galactic plane. 
The {\it Planck} survey covered the entire sky in 9 bands with wavelengths from 30 GHz to 857 GHz (1 cm to 350 $\mu$m) \citep{tauber2010a, Planck2011-1.1}. 
The Galactic L-band Feed Array (GALFA) H~I survey covered the sky between declinations 
0$^\circ$--40$^\circ$ \citep{peek11}.
The clouds are at high galactic latitude and can be assumed to lie at a typical distance of $\sim 100$ pc based on the scale-height (150 pc) of cold, atomic gas in the Milky Way \citep{kalberla09}. 

To measure the radio absorption by H~I and OH toward the clouds, we identified the continuum sources from the NRAO VLA All Sky Survey \citep{condon98} near each cloud in our sample. Each source was
given a priority based on its radio continuum flux and location relative to the peak in the
`dark gas' map. Sources with flux density at 1400 MHz greater than 300 mJy were expected to yield
easy measurements of 21-cm absorption, and sources fainter than 100 mJy were not feasible.
We discarded all sources fainter than 100 mJy, and sources fainter than 300 mJy that were not located
well within the outer boundary of the cloud at the $10^{20}$ cm$^{-2}$ column density level.
All remaining sources were observed for 21-cm absorption. 
The brightest sources were also observed, with longer integration time, for OH 1665 and 1667 MHz absorption. The OH absorption observations were intended to test the hypothesis that there is
significant molecular gas associated with the dark gas.
Table~\ref{obstab} lists the radio sources, their radio fluxes, and the amount of dark gas based
on the Paper 1 maps.
The sources observed only for 21-cm line absorption are designated in the second column with `HI', while
the ones with longer integrations on the 1667/1667 MHz absorption are designated `OH'.

The initial observations occurred in 2014 ?? for the radio sources toward the cloud G254+63.
Absorption lines at 21-cm were detected toward many sources, while OH absorption was not.
Based on those results, our more complete survey of the clouds in 2015 focussed on 21-cm absorption
and had generally shorter integrations except toward a select few sources (listed in Table~\ref{obstab}). 

We also observed the peaks of the `dark gas' emission to search for OH emission. These observations were different from the others described in this paper because they were toward positions with no radio continuum source. \citet{allen15} recently found emission from diffuse OH associated
with `dark gas' that is more closely 
associated with atomic gas and has no CO counterpart. Those observations were close
to the galactic plane; our goal was to test whether such emission was present toward
the `dark gas' peaks of our well characterized sample of high latitude clouds.
No OH emission lines were detected, so these observations will not be analyzed in detail. 
Upper limits on the column density require assuming the OH excitation temperature above the cosmic microwave
background, and the uncertainties in excitation are large enough that the results would be of no help
in our quest to determine the nature of the dark gas in the clouds.

\def\tnm{\tablenotemark}
\begin{deluxetable}{ccccccccl}
\tabletypesize{\scriptsize} 
\tablecolumns{9}
\tablecaption{Selected radio sources behind interstellar clouds}\label{obstab} 
\tablehead{
\colhead{Source} &   \colhead{Type\tnm{a}} &    \colhead{RA} & \colhead{Dec}    & \colhead{$F_{1400}$ \tnm{b}} & 
\colhead{$\tau_{353}$\tnm{c}} & \colhead{$N({\rm H~I})$\tnm{d}} & \colhead{N({\rm dark})\tnm{e}} & 
\colhead{Note\tnm{f}}\\
& & (J2000) & (J2000) & (mJy) & ($10^{-5}$) & {($10^{20}$ cm$^{-2}$)} & {($10^{20}$ cm$^{-2}$)}
}
\startdata
\sidehead{G94-36}
 2  &  HI  &  23 01 56.60 & +21 07 53.3 & $\phn226\pm 8\phn$  & 1.3 & 5.8 & \phd0.5 & just outside main core\\
 3  &  OH  &  23 02 15.60 & +21 21 42.6 & $\phn392\pm 12$     & 2.5 & 5.6 & \phd4.7  & {\bf at edge of core}\\
 5  &  HI  &  23 01 23.06 & +20 58 56.9 & $\phn103\pm 4\phn$  & 1.0 & 5.9 & -0.2 & none\\
 6  &  HI  &  23 09 23.15 & +20 54 22.6 & $\phn152\pm 5\phn$  & 1.3 & 5.7 & \phd1.2  & moderate, diffuse\\
 7  &  HI  &  23 08 11.63 & +20 08 42.3 & $\phn188\pm 6\phn$  & 1.3 & 5.0 & \phd1.5  & moderate, diffuse\\
 8  &  HI  &  23 00 39.87 & +20 44 21.6 & $\phn256\pm 9\phn$  & 0.5 & 4.4 & -0.5 & none\\
\sidehead{G104-39}
 2  &  HI  &  23 47 55.83 & +22 17 02.5 & $\phn418\pm 13$     & 0.5 & 3.4 & \phd0.7  & diffuse\\
 3  &  OH  &  23 47 49.36 & +22 00 16.4 & $\phn518\pm 16$     & 1.1 & 4.2 & \phd2.4  & {\bf at edge of core}\\
 4  &  HI  &  23 45 16.18 & +21 51 41.9 & $\phn199\pm 7\phn$  & 0.8 & 4.4 & \phd0.6  & diffuse\\
 5  &  HI  &  23 52 40.34 & +21 57 35.6 & $\phn156\pm 5\phn$  & 0.6 & 3.6 & \phd0.9  & diffuse\\
 7  &  HI  &  23 52 20.72 & +21 10 30.3 & $\phn274\pm 10$     & 0.5 & 3.5 & \phd0.1  & none\\
\sidehead{G108-53}
 1  &  OH  &  00 16 27.25 & +09 29 23.6 & $\phn127\pm 4\phn$  & 1.9 & 7.0 & \phd4.1  & {\bf behind core}\\
 2  &  HI  &  00 18 55.26 & +09 40 05.9 & $\phn489\pm 15$     & 0.9 & 5.8 & \phd0.7  & just outside main core\\
 3  &  HI  &  00 14 44.86 & +09 24 29.4 & $\phn100\pm 3\phn$  & 1.4 & 7.8 & \phd0.8  & just outside main core\\
 5  &  HI  &  00 14 19.89 & +08 54 02.2 & $\phn324\pm 12$     & 1.0 & 6.3 & -0.2 & none\\
 6  &  HI  &  00 19 12.36 & +08 40 53.8 & $\phn494\pm 17$     & 1.1 & 6.6 & \phd0.9  & moderate, diffuse\\
 7  &  HI  &  00 17 38.31 & +08 27 46.8 & $\phn211\pm 8\phn$  & 1.3 & 6.1 & \phd1.9  & {\bf behind small core}\\
 8  &  HI  &  00 17 15.73 & +08 26 06.7 & $\phn219\pm 8\phn$  & 1.0 & 6.0 & \phd1.0  & moderate, diffuse\\
 10 &  HI &   00 15 08.4 &  +08 28 03.8 & $\phn384\pm 12$     & 1.1 & 5.8 & \phd0.9  & moderate, diffuse\\
\sidehead{G138-52}
 1  &  HI  &  09 08 45.42 & +09 43 21.6 & $\phn110\pm 3$      & 0.7 & 6.8 & -0.2 & none\\
 2  &  HI  &  09 11 22.12 & +09 16 55.7 & $\phn115\pm 4$      & 0.5 & 5.8 & \phd0.1  & none\\
 4  &  HI  &  09 06 11.61 & +08 44 33.5 & $\phn108\pm 4$      & 0.5 & 5.5 & \phd0.1  & none\\
 6  &  HI  &  09 12 31.32 & +09 13 24.5 & $\phn257\pm 8$      & 0.5 & 5.5 & \phd0.1  & none\\
 9  &  HI  &  09 03 40.45 & +10 03 22.3 & $\phn108\pm 3$      & 0.6 & 5.5 & \phd0.3  & none\\
\sidehead{G198+32}
 1  &  HI  &  08 27 33.76 & +26 37 16.5 & $\phn102\pm 3$      & 0.8 & 3.9 & \phd1.9  & {\bf at edge of core}\\
 2  &  OH  &  08 25 56.92 & +26 43 57.8 & $\phn223\pm 7$      & 0.6 & 3.3 & \phd2.0  & {\bf at edge of core}\\
 3  &  HI  &  08 25 47.36 & +27 04 21.7 & $\phn108\pm 3$      & 0.2 & 2.8 & -0.4 & none\\
 6  &  HI  &  08 29 48.12 & +25 04 10.3 & $\phn126\pm 4$      & 0.3 & 3.3 & \phd0.5  & diffuse\\
 9  &  HI  &  08 28 59.56 & +24 54 00.9 & $\phn219\pm 7$      & 0.3 & 3.1 & \phd0.4  & none\\
\sidehead{G221+35}
 1  &  HI  &  01 30 43.15 & +10 16 30.3 & $\phn101\pm 4\phn$  & 0.6 & 5.2 & \phd0.1 & just outside main core\\
 2  &  HI  &  01 29 04.27 & +10 45 28.9 & $\phn111\pm 3\phn$  & 0.5 & 5.4 & -0.1 & none\\
 4  &  HI  &  01 26 10.02 & +10 14 01.2 & $\phn670\pm 23$     & 0.5 & 5.0 & \phd0.4 & diffuse\\
 5  &  HI  &  01 29 09.07 & +09 18 21.7 & $\phn130\pm 4\phn$  & 0.5 & 5.2 & \phd0.1 & none\\
 6  &  HI  &  01 33 38.98 & +10 19 43.7 & $\phn391\pm 12$     & 0.4 & 4.7 & -0.3 & none\\
 7  &  HI  &  01 27 23.20 & +09 14 31.9 & $\phn134\pm 4\phn$  & 0.5 & 5.1 & \phd0.6 & moderate, diffuse\\
 8  &  HI  &  01 29 22.11 & +11 19 49.8 & $\phn181\pm 3\phn$  & 0.7 & 3.8 & \phd3.5 & {\bf behind core} \\
 9  &  HI  &  01 34 16.62 & +10 25 42.7 & $\phn129\pm 4\phn$  & 0.4 & 3.3 & \phd1.5 & just outside core\\
 10 &  HI &   01 30 11.9 &  +11 28 53.8 & $\phn233\pm 9\phn$  & 0.4 & 3.5 & \phd0.5 & just outside core\\
\tablebreak
\sidehead{G236+39}
 1  &  HI  &  09 47 44.60 & +00 04 37.2 & $\phn936\pm 33$     & 0.8 & 5.0 & \phd1.7 & {\bf behind small core} \\
 2  &  HI  &  09 43 21.09 & +01 03 38.8 & $\phn169\pm 6\phn$  & 0.7 & 6.0 & -0.1 & none\\
 3  &  HI  &  09 43 16.93 & +01 02 43.0 & $\phn234\pm 8\phn$  & 0.7 & 6.1 & -0.1 & none\\
 4  &  HI  &  09 43 19.16 & -00 04 22.3 & $1190\pm 40$    & 0.4 & 3.6 & -0.1 & none \\
 6  &  HI  &  09 47 15.89 & -00 42 45.1 & $\phn230\pm 7\phn$  & 0.4 & ... & ...  & ...\\
 9  &  HI  &  09 51 36.11 & +00 53 15.6 & $\phn356\pm 11$     & 0.6 & 3.7 & \phd1.6 & {\bf behind small core}\\
\sidehead{G254+63}
 5  & OH   &  11 35 25.04 & +06 54 40.7 & $\phn273\pm 9\phn$  & 0.7 & 3.4 & \phd2.6 & {\bf just at edge of core}\\
 7  & OH   &  11 30 45.69 & +07 26 21.3 & $\phn445\pm 13$     & 0.6 & 4.1 & \phd1.3 & {\bf just at edge of core}\\
 11 & OH   &  11 25 13.21 & +07 06 34.0 & $\phn112\pm 3\phn$  & 0.9 & 5.3 & \phd1.9 & {\bf just at edge of core}\\
 12 & OH   &  11 23 09.10 & +05 30 20.3 & $1720\pm 17$    & 0.3 & 2.6 & \phd0.6 & none \\
 14 & OH   &  11 16 17.79 & +05 43 52.6 & $\phn335\pm 10$     & 0.7 & 6.5 & -0.1 & none \\
\enddata
\tablenotetext{a}{Type of observation. "HI" means a short observation for H~I 1420 MHz absorption. "OH" means a longer (typically 90-minute) observation for OH 1667 MHz absorption.}
\tablenotetext{b}{1400 MHz Flux of the radio continuum source in Jy, from the NRAO VLA Sky Survey}
\tablenotetext{c}{Optical depth of dust at 353 GHz from the {\it Planck} survey.}
\tablenotetext{d}{Column density, $N_{\rm naive}$, of atomic hydrogen from the GALFA survey assuming the line 21-cm is optically thin.}
\tablenotetext{e}{Column density of `dark gas' at the source location, using methods from Paper 1.}
\tablenotetext{f}{Qualitative description of the location of the radio source relative to the spatial distribution of `dark gas' from the {\it Planck}-Arecibo image.}
\end{deluxetable}

\def\etra{
\def\tnm{\tablenotemark}
\begin{deluxetable}{ccccccc}
\tabletypesize{\scriptsize} 
\tablecolumns{9}
\tablecaption{Column density fits}\label{fittab} 
\tablehead{
\colhead{Source} &  \colhead{$\tau_{\rm CNM}$}  & \colhead{$T_s$} & \colhead{$\sigma_{T_s}$} & 
\colhead{$N_{\rm CNM}$} & \colhead{$N_{\rm WNM}$} & \colhead{$N_{\rm naive}$} \\
& Peak & (K) & {($10^{20}$ cm$^{-2}$)} & {($10^{20}$ cm$^{-2}$)} & {($10^{20}$ cm$^{-2}$)}
}
\startdata
 5  & 1.71 & 16.5 & 3.6 & 1.01 & 2.36 & 2.72\\
 7  & 0.44 & 43.8 & 2.5 & 1.38 & 2.25 & 3.79\\
 11 & 0.75 & 10.7 & 5   & 0.25 & 3.80 & 3.77\\
 12 & 0.21 & 39.4 & 2.2 & 0.32 & 2.13 & 2.46 \\
 14 & 0.42 & 64.2 & 5.2 & 1.70 & 3.52 & 4.88 \\
\enddata
\tablenotetext{a}{Fits in this table are for $\mathcal{F}=0.5$}
\end{deluxetable}
}

\subsection{Observation execution}
All observations for this paper were performed on the Arecibo Observatory 305-m radio telescope using the L-wide feed, which was tuned to the 21-cm line of H~I at 1420 MHz or the OH lines at 1665 and 1667 MHz.
The gain of the L-wide receiver is 10.5 K/Jy at at 1415 MHz and 9.9 K/Jy at 1666 MHz in the receiver's
`post-2003' configuration.

The pilot project on G254+63 was executed in 2014 August as project A2902. The remainder of the sample was observed in 2015 July as project A2985.
The observing sequence and data reduction were the same as used for the Millennium Survey 
\citep{heiles03}.



\begin{figure}[!htb]
\includegraphics[scale=.4]{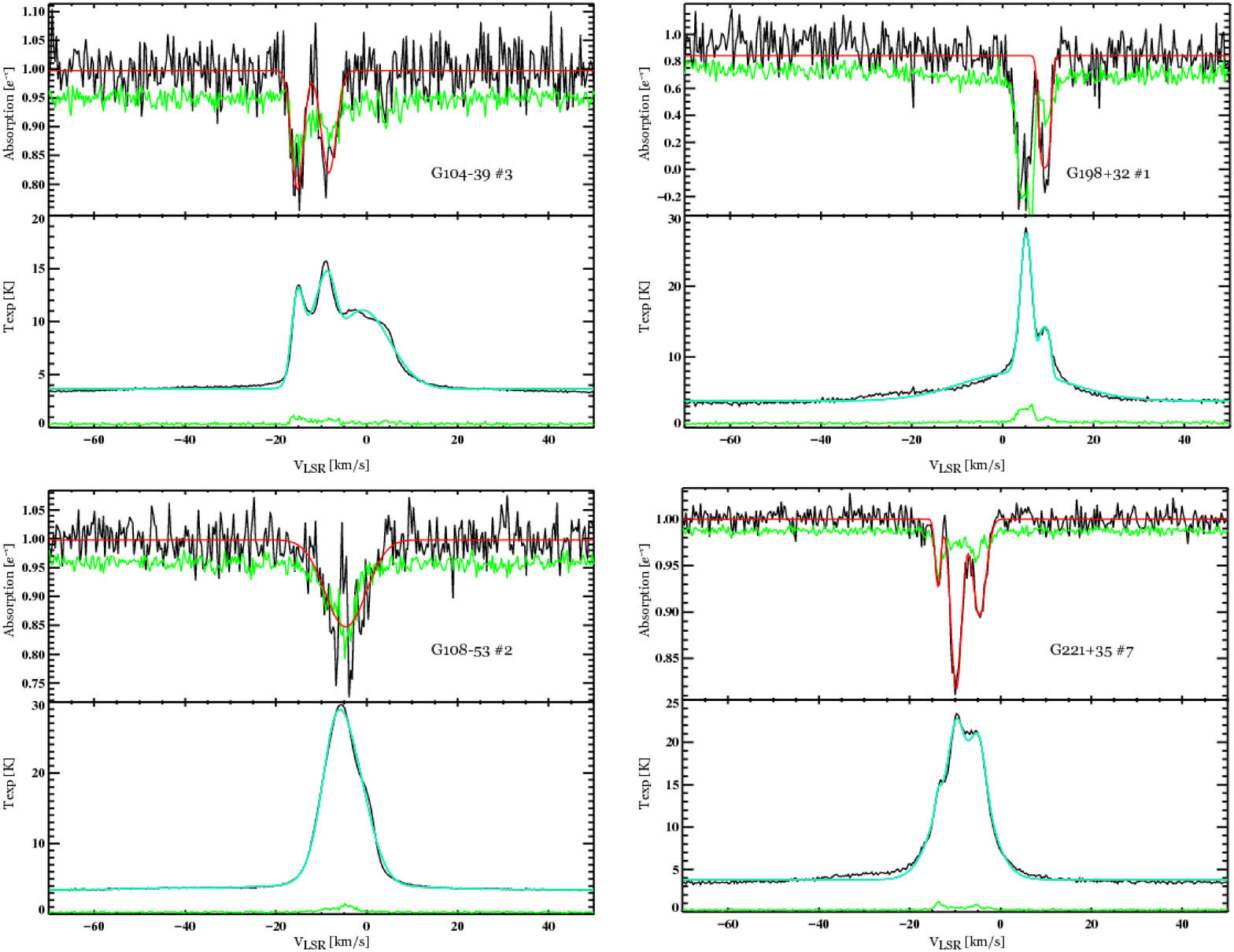}
\caption[absplots]{
H~I 21-cm emission and absorption spectra for 4 radio sources from our survey (one source from each cloud with statistically 
significantly detected cold atomic gas). 
For each source, there is a two-panel plot. The top panel is the absorption spectrum ($e^{-\tau}$) toward the radio source; the lower
panel is the emission spectrum at the location of the radio source ($T_{\rm exp}$). The observed data are in 
black, the fit to significant absorption components in red, the fit to the emission spectrum is in cyan, and the uncertainty in green.
The absorption components where the black curve dips below green are the believable ones, while the apparent absorptions that are within the
uncertainties are likely due to fluctuations in the emission spectrum on the scale of the Arecibo beam.
\label{absplots}}
\end{figure}

Figure~\ref{absplots} shows four of the H~I emission and absorption spectra obtained from this survey. 
The uncertainty absorption spectra comprises two components: noise due to the sensitivity of the receiver for measuring the brightness of the radio continuum source, and `pseudo-noise' due to incomplete subtraction of the 21-cm emission toward the radio source due to fluctuations in
the emission between the `on' and `off' source beams. The noise increases for fainter sources
in a predictable manner and is approximately independent of frequency. 
The `pseudo-noise' depends on the amount of structure in the 21-cm emission on angular scales
comparable to the beam ($3^\prime$), and it has a frequency-dependence similar to that of the
21-cm emission. The methods for estimating the noise components and fitting the signal are described
in the next section of this paper.

We see no OH absorption toward any of the radio sources. Upper limits to the optical depths in 1165 and 1667 MHz OH lines are
typically less than 0.01. The column density limits were derived from the combination of 1665 and 1667 OH profiles, assuming 
the excitation temperature of the hyperfine levels is $T_x = 5$ K and assuming that the OH central velocity and width are 
identical to the HI CNM components. 
The upper limits to the OH column densities are $N({\rm OH}) < 3\times 10^{13}$ cm$^{-2}$.
For comparison the `dark gas' column densities are typically $2\times 10^{-20}$, so the upper limit to the abundance of OH
in the `dark gas' is [OH]/[H~I]$< 1.5\times 10^{-7}$.
For comparison, for diffuse and translucent clouds radio observations have found an abundance ratio (for an assumed $T_x=5$ K)
[OH]/[H~I]=$4\times 10^{-8}$, with these values being nominally lower limits \citep{lisztlucas}. 
The upper limits from our observations are higher than these lower limits, so our observations are not very restrictive.
If the `dark gas' were all H$_2$, with abundance [OH]/[H$_2$]=$8\times 10^{-8}$, then we expect a typical OH column density of $1.6\times 10^{13}$ cm$^{-2}$. To detect such
lines at 3$\sigma$ would require 5 times greater signal-to-noise, which is beyond the limits of the techniques
used in this project because we found the uncertainties are dominated by systematics in the variations of the H~I emission
on the angular scale of the Arecibo beam. 

\def\extra{
{\bf CARL, IN THIS SECTION WE COULD DOCUMENT THE DEEP EMISSION STARES ON OH. I THINK THAT DERIVING THE LIMITS COULD BE DONE FROM THE AVERAGE SPECTRUM BY ASSUMING THE SAME LINE CENTER AND WIDTH AS FOR THE CNM COMPONENTS, LIKE YOU DID FOR THE ABSORPTION. THEN GET UPPER LIMIT TO THE LINE INTEGRAL AND THEN USE TX=5 TO CONVERT TO COLUMN DENSITY. I WOULD LIKE TO KNOW WHAT THESE COME OUT TO BE AND I THINK THEY MAY BE WORTH INCLUDING.}
}

\pagebreak
\section{How much optically-thick gas is present?}

\subsection{Deriving the gas temperature and column density}

We follow the method described in \citet{heiles03} to simultaneously determine the absorption spectrum toward the radio source and the `expected' emission spectrum that would obtain if the radio source were not present. The procedure solves for the spatial gradients of the emission between the radio source and the adjacent positions used to measure the `off' source emission, and interpolates the emission spectra to yield the `expected' emission spectra at the location of the source. Taking care in performing this interpolation is important, because small variations in the column density of the cloud between the `on' and `off' source observations produce fluctuations that mask or confuse the absorption spectrum; furthermore, to interpret the absorption spectrum, we need to compare to the emission spectrum at the same location to the extent possible. The fundamental limitation of the interpolation is the size of the the telescope beam. While the absorption samples a pencil-thin beam toward the radio source, the emission spectrum averages over a $3'$ beam. We use the variation among the multiple `off' beams to assess both the spatial 
first and second derivatives of the H~I intensity
and their uncertainties, allowing us to derive an uncertainty estimate for the absorption and `expected' emission spectra. 

The 21-cm emission is assumed to arise from multiple components at different temperatures. For each component, the 21-cm emission in absence of absorption by other components is 
\citep[following][]{heiles03}
\begin{equation}
T_{\rm exp}(v) = T_s \left( 1 - e^{-\tau(v)} \right)
\end{equation}
where $T_s$ is the spin temperature of the H~I hyperfine state, and $\tau$ is the optical depth.
The optical depth is related to the column density, $N(v)$, as
\begin{equation}
\tau(\nu) =  N(v)/C T_s
\label{eq:ncnm}
\end{equation}
where $C=1.823\times 10^{18}$ cm$^{-2}$~K$^{-1}$~s.
Emission arises from both the warm, neutral, medium (WNM) and the cold, neutral, medium (CNM).
The WNM is warm enough ($T_{s,{\rm WNM}}>1000$ K) that its optical depth is small, 
$\tau_{\rm WNM}\ll 1$, and its emission spectrum
\begin{equation}
T_{\rm exp,{\rm WNM}} = T_{s,{\rm WNM}} \times \tau_{WNM}(v) =   N_{\rm WNM}(v)/C .
\label{eq:nwnm}
\end{equation}


The cold gas provides both narrow emission components and absorption, both of the background radio source and part of the warm gas. For the isolated clouds in this paper, there is generally a dominant 
cold component on the line of sight,
and the cloud can be readily seen in emission at that velocity. 
The emission from the CNM is
\begin{equation}
T_{\rm exp, {\rm CNM}} = T_{s,{\rm CNM}} \left(1 - e^{-\tau_{\rm CNM}(v)} \right).
\label{eq:tcnm}
\end{equation}
The CNM is assumed to be in a spatially compact structure (both in the sky and along the line of sight), while the WNM may extend along a longer path length. A fraction $\mathcal{F}$ of the WNM lies in front of the cloud and is unabsorbed, while the
background WNM is absorbed by the CNM. 
The total expected spectrum is then
\begin{equation}
T_{\rm exp} = T_{\rm exp,{\rm WNM}} \left[ \mathcal{F} + (1-\mathcal{F}) 
e^{-\tau_{\rm CNM}(v)} 
\right]
+T_{s,{\rm CNM}} \left(1 - e^{-\tau_{\rm CNM}(v)} \right).
\label{eq:texp}\end{equation}

For the line of sight toward the radio source, the brightness difference on- and off-source,
$\Delta T$, directly yields the absorption spectrum of the CNM,
\begin{equation}
\Delta T / T_{src} = e^{-\tau_{\rm CNM}(v)}
\label{eq:taucnm}
\end{equation}
where $T_{src}$ is the brightness temperature of the radio source. 

We fit equation~\ref{eq:taucnm} to determine the Gaussian components of the cold gas absorption 
$\tau_{\rm CNM}(v)$.
Then we fit equation~\ref{eq:texp}, for an assumed value of $\mathcal{F}$, to 
to solve for the spin temperatures $T_{s,{\rm CNM}}$ of the cold gas and a sum of Gaussian
emission components of the warm gas, $N_{\rm WNM}(v)$. The column density of cold gas
was finally determined from the Gaussian fit amplitudes of the CNM components and spin temperature:
$N_{CNM}=C\times \tau_{CNM} \times T_{s,{\rm CNM}}$.

\subsection{Effect of foreground WNM}
The fraction of foreground WNM is generally not known. The extreme case of having all
the WNM in front of the cloud ($\mathcal{F}=1$) is  unlikely, given that the WNM has a much
larger filling factor and scale-height than the CNM \citep{cox05,malhotra95}. 
The extreme case of having all the WNM behind the cloud ($\mathcal{F}=0$) is also unlikely because it would mean the cloud is so close to the Sun that
it would subtend a very large angle. Because our survey covers several clouds that are typical of the
ISM as seen in all-sky 21-cm surveys, the most likely case is an intermediate value 
$\mathcal{F}\simeq 0.5$, for the cloud ensemble average, because the scale-heights of the WNM and
CNM are comparable \citep{dickey09}. 
Apart from the arguments about relative scale height, the components of WNM and CNM that overlap in
velocity are likely to be related to each other as part of the same spatial region. Again, statistically,
this argues for $\mathcal{F}\simeq 0.5$ on an ensemble average. 
Extreme values of $\mathcal{F}$ could
occur if the warm gas is associated with the cloud but is located asymmetrically with respect to it. 
For example, if the cloud is being shocked or radiatively heated, and energy input is on the far-side
of the cloud, $\mathcal{F}=0$; if the energy input is on the near-side of the cloud,
$\mathcal{F}=1$.

\begin{figure}[!htb]
\includegraphics[scale=.8]{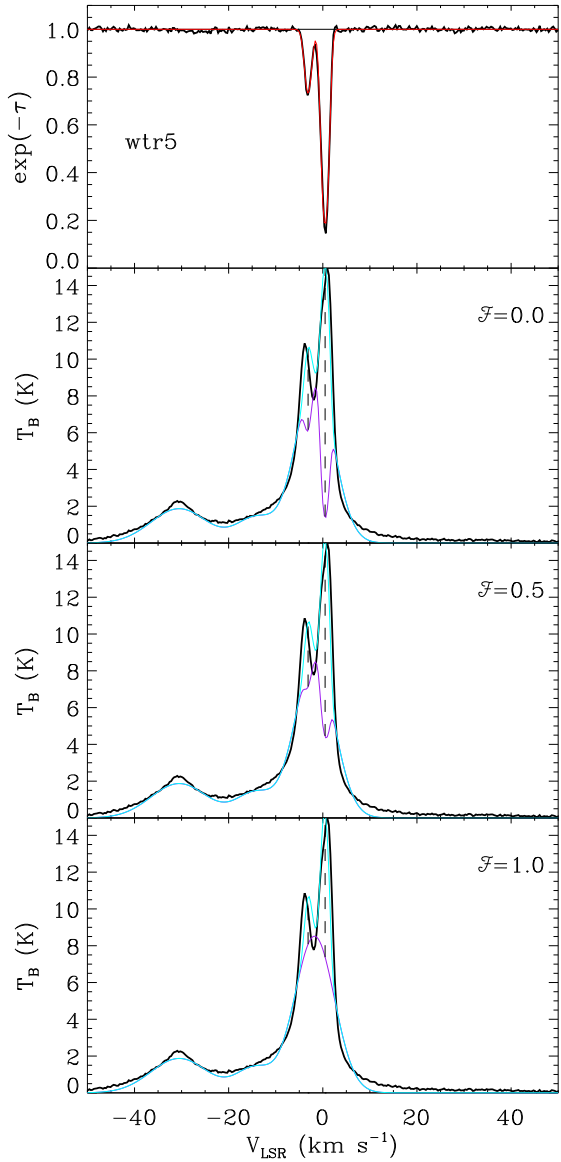}
\caption[plotfit3]{
H~I 21-cm emission and absorption towards radio source G254+63\#5, showing the effects of 
assuming different fractions of foreground warm gas. 
The top panel shows the absorption spectrum, with the data in black and the Gaussian-fitted absorption
spectrum in red (as in Fig.~\ref{absplots}).
The lower three panels each show the expected emission spectrum (thick black curves), $T_{\rm exp}$,
together with the fitted warm gas contribution (thin purple curves) to the emission spectrum,
$T_{\rm exp, WNM}$, for an assumed fraction, $\mathcal{F}$ (labeled) of warm gas in front of the cold cloud.
The summed warm and cold gas for each fit are shown in cyan (as in Fig.~\ref{absplots}).
The dashed lines show the location and amplitude 
of the CNM contributions both the emission and absorption spectra. 
\label{plotfit3}}
\end{figure}

To illustrate the technique, we show in Figure~\ref{plotfit3} the observations and fit for a
source behind G254+63. The fit to the CNM absorption (top panel) is straightforward because the profile is well-fitted with 2 Gaussian components and we need not assume anything about $\mathcal{F}$.
The fit to the emission
involves more parameters and is not perfect. We strive for simplicity in the fits, utilizing only
a small number of velocity components, and this approach can never match the observed, structured,
high-signal-to-noise profiles. The value of the fits with few components is that the properties of the gas can be relatively
reliably estimated; including more components would better fit the profiles but yield less 
reliable spin temperature fits, because it would be driven by extra Gaussians included to
account for the non-Gaussianity of the velocity profile.
For the fits in Figure~\ref{plotfit3}, there were only 3 WNM Gaussian components, 
but 2 of them are at -30.5 and -14.3 km~s$^{-1}$, putting them out of the `action' for determining the column density of the 
cold H~I at the velocity of the cloud. 
The shape of the fitted WNM emission near the cloud velocity is complicated for 
$\mathcal{F}<1$, despite being due
to just a single Gaussian emission component, because the WNM is absorbed 
by the cold gas in the cloud. 
The spin temperature of the CNM is determined by the amplitude of the CNM contribution to the emission spectrum, so it depends on the way the WNM and CNM are differentiated.
The fits in Figure~\ref{plotfit3} show what happens if we assume
$\mathcal{F}=$0, 0.5, and 1. 
If none of the WNM is in front of the cloud ($\mathcal{F}=0$), 
then the WNM experiences the full absorption by the CNM;
the fit to the emission spectrum then requires relatively more CNM contribution, and the inferred
spin temperature of the CNM is higher. In contrast, if the WNM is in front of the CNM ($\mathcal{F}=1$),
it experiences no absorption, and less emission from the CNM is required to match the observed emission
spectrum.
This effect can be seen graphically in the Figures by the lengths of the dashed lines, which are proportional to the spin temperature of the CNM. For this source, the derived CNM spin temperatures 
are 20.2, 16.5, or 13 K, for $\mathcal{F}$ of 0, 0.5, or 1, respectively. 

After solving for the WNM and CNM parameters, their column densities are determined by
for each Gaussian component using equation~\ref{eq:nwnm} for the WNM and~\ref{eq:ncnm} for
the CNM. The properties of the WNM do not really depend upon the fit details, but the amount
of CNM does depend on $\mathcal{F}$ because the optical depth is inversely proportional to the
spin temperature and the derivation of spin temperature depends on the relative location of the
cloud with respect to the WNM. For the same radio source shown in Figure~\ref{plotfit3}, 
we find that for $\mathcal{F}$=0, 0.5, or 1, we find
$N_{\rm CNM}$/10$^{20}$ cm$^{-2}$ of 1.2, 1.0, and 0.8, respectively. 
Thus even with the complete
range of possible locations of the the WNM, the effect on the cold gas column density is
$\pm 20$\%.

Table~\ref{fittab} lists the column densities determined from the fits for the sources with statistically-significantly detected column densities of cold gas.

\def\extra{
{\bf THERE IS A BIT OF A PROBLEM HERE BECAUSE IN THE END, CARL DIDN'T BELIEVE THE SPECTRUM OF THIS SOURCE YIELDED A COLUMN DENSITY WORTH REPORTING. THERE ARE SO FEW SOURCES WITH COLUMN DENSITIES THAT MAYBE THIS SECTION ABOUT FOREGROUND GAS CAN BE REMOVED?}
}

\def\tnm{\tablenotemark}
\begin{deluxetable}{lccccc}
\tabletypesize{\scriptsize} 
\tablecolumns{5}
\tablecaption{Column density fits\tnm{a}\label{fittab}} 
\tablehead{
\colhead{Source} & \colhead{$N_{\rm CNM}$} & \colhead{$N_{\rm WNM}$} &  \colhead{$N_{\rm true}$} & \colhead{$N_{\rm naive}$} 
\\
                  & {($10^{20}$ cm$^{-2}$)} & {($10^{20}$ cm$^{-2}$)} & {($10^{20}$ cm$^{-2}$)} & {($10^{20}$ cm$^{-2}$)} 
}
\startdata
 G104-39 \#2 & $0.37 \pm 0.07$ & $2.55$ & $2.92 \pm 0.07$ & $2.76$ \\ 
 G104-39 \#3 & $1.07 \pm 0.18$ & $2.68$ & $3.74 \pm 0.18$ & $3.45$ \\ 
 G108-53 \#2 & $4.41 \pm 0.46$ & $1.42$ & $5.83 \pm 0.46$ & $5.35$ \\ 
 G108-53 \#6 & $2.50 \pm 0.46$ & $2.72$ & $5.22 \pm 0.45$ & $4.54$ \\ 
 G108-53 \#8 & $4.06 \pm 1.42$ & $2.74$ & $6.80 \pm 1.42$ & $4.71$ \\ 
G108-53 \#10 & $3.06 \pm 0.52$ & $2.43$ & $5.49 \pm 0.52$ & $5.06$ \\ 
 G138-52 \#6 & $0.88 \pm 0.36$ & $3.75$ & $4.63 \pm 0.36$ & $4.23$ \\ 
 G198+32 \#1 & $2.12 \pm 0.62$ & $4.02$ & $6.14 \pm 0.62$ & $3.71$ \\ 
 G198+32 \#2 & $0.89 \pm 0.33$ & $2.99$ & $3.89 \pm 0.33$ & $3.17$ \\ 
 G221+35 \#7 & $0.95 \pm 0.08$ & $3.53$ & $4.48 \pm 0.08$ & $4.14$ \\ 
 \enddata
\tablenotetext{a}{Fits in this table are for $\mathcal{F}=0.5$}
\end{deluxetable}

\def\extra{
\def\tnm{\tablenotemark}
\begin{table}
\caption[]{OH Emission Observation Summary}\label{ohobstab} 
\begin{flushleft} 
\begin{tabular}{lcccc}
\hline\hline
Cloud\tnm{a}   & IREX & Time &    RA         & Dec   \\ \hline
G198+32    & Highest  & ??    & 08:27:09.974 & +26:15:30.19 \\
G236+39    & Low      & ??    & 09:45:45.984 & +00:45:30.0 \\
G94-36     & Highest  & ??    & 23:04:42.080 &  +21:00:30.15 \\
G104-39    & Low      & ??    & 23:53:26.086 & +22:32:30.16 \\
G108-53    & Moderate & ??    & 00:16:33.916 & +09:52:30.07 \\
\end{tabular}
\end{flushleft} 
\end{table}  
}

\clearpage

\section{Upper limit to Column Density of Optically-thick H~I}

We set upper limits on the column density of cold, optically-thick H~I for lines of sight with no detected absorption using the following simple prescription.
For a single, isothermal interstellar cloud, the total H~I column density depends on the optical depth profile
\begin{equation}
N_{\rm true} = C T_s \int \tau dv
\end{equation}
while the na\"ive column density depends on the line integral
\begin{equation}
N_{\rm naive} = \int T_B dv = C T_s \int (1-e^{-\tau}) dv.
\end{equation}
Using a power-series for the exponential, and 
assuming Gaussian optical depth profiles with peak value $\tau_0$, we find
\begin{equation}
\frac{\Delta N}{N_{\rm naive}} \equiv \frac{N_{\rm true}-N_{\rm naive}}{N_{\rm true}}
= \sum_{n=2}^{\infty}  \frac{(-1)^n}{n! n^{0.5}} \tau_0^{n-1}
\label{eq:deln}
\end{equation}
For optical depths less than unity, a second-order expansion is likely sufficient; then
\begin{equation}
\label{eq:deln}
\frac{\Delta N}{N}  \simeq 0.4 \tau_0.
\end{equation} 
Therefore, if we have an upper limit $\tau_0 < 0.5$, we find that the deviation between 
naive and true optical depth is less than 20\%.
The upper limits on the cold H~I are calculated using the upper limits on the optical depth 
($\tau_0 < 0.3$ for radio sources brighter than 300 mJy and $\tau_0<0.7$ for radio sources between 100--300 mJy),
and equation~\ref{eq:deln}.

\section{Comparison between `Dark Gas' and Optically-thick H~I}

\begin{figure}[!tbh]
\includegraphics[scale=1]{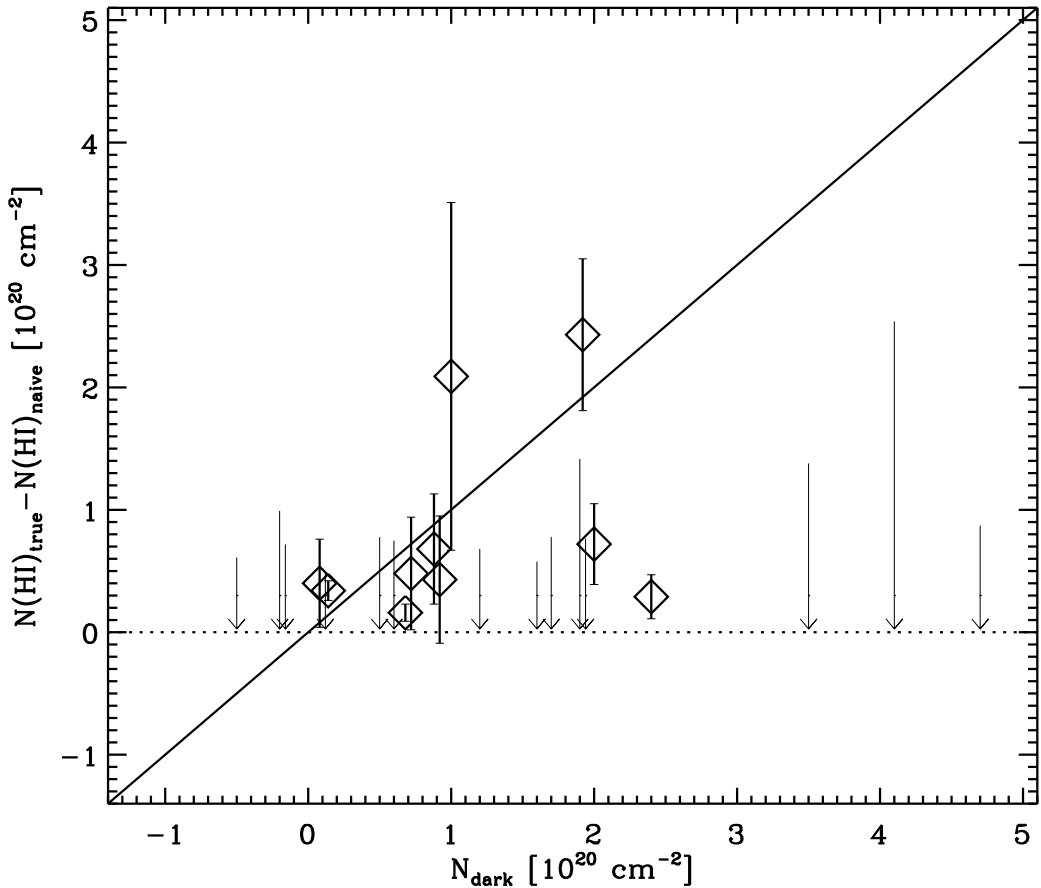}
\caption[ndelndark]{
The column density of cold H~I determined from the 21-cm absorption observations, compared to the column density of ``dark''
gas determined from the {\it Planck} dust and 21-cm emission observations. Open diamonds with error bars show the results
for lines of sight where the 21-cm absorption was detected. Grey downward arrows show the upper limits for the lines of sight
with no detected H~I absorption. The horizontal dotted line just shows the level of zero optically-thick gas, and the diagonal line
just shows the unity relation. If all the `dark' gas were due to optically-thick atomic gas, we would expect the points to line along
the diagonal line.
\label{ndelndark}}
\end{figure}

Now that the amount of cold, optically-thick H~I has been assessed for the lines of sight toward high-latitude clouds,
we can compare to the amount of `dark' gas that had been inferred from the comparison of {\it Planck} dust emission to
21-cm emission, from Paper 1. Figure~\ref{ndelndark} directly compares the two quantities. 
There are only upper limits to the 21-cm absorption for most of the points, including those toward the 
highest concentrations of dark gas. This is due to the cloud cores spanning less solid angle, making
the number of bright radio sources smaller there. The fainter radio sources, which we did all observe, sometimes
with relatively long integration times, at least allowed upper limits that appear to rule out the 
hypothesis that the dark gas is optically-thick atomic gas. Considering only the lines of sight with
detections, there is no clear trend. 
It does appear that the two lines of sight with the highest amount of optically-thick atomic gas do indeed have `dark' gas; furthermore, the {\it amount} of optically-thick atomic gas appears consistent with the amount of 
dark gas for these two points. However, there are two other detections, with comparable or larger amounts of dark gas, that have far less optically-thick atomic gas than would be required to explain the entire dark gas content.

We can understand these results as follows. First, when there is cold, optically-thick atomic gas, there should be
dust mixed with it. Therefore, points in Figure~\ref{ndelndark} above zero vertically, should move along the diagonal line, just due to the dust associated with the cold gas. There should {\it not} be points in the upper-left half of the diagram: such points would have cold atomic gas without associated dust. This explains why
Figure~\ref{ndelndark} is populated in the lower-right portion only. 
The presence of points that are clearly {\it below} the diagonal line---especially those toward the right-hand 
side---shows that there is more dark gas than can be explained by cold atomic gas.

\section{Conclusions}

Our results indicate two likely origins for the dark gas. 
There is a small amount that arises from cold atomic gas, on at least two lines of sight.
However, if the dark gas column density estimates are correct,
the bulk of the `day' gas associated with the dust must be
in a separate form, the most likely choice is molecular gas, which remains
the default hypothesis \citep{wolfire10}. 
There remains the possibility that dark gas estimates based
on dust column density
are incorrect, which would mean that the locations of `apparent' dark gas may instead be regions of enhanced 
submm dust emissivity, as discussed in Paper 1. We will attempt to quantify that possibility in future work.


\acknowledgements  
The Arecibo
Observatory is operated by SRI International under a cooperative
agreement with the National Science Foundation (AST-1100968), and in
alliance with Ana G. M\'endez-Universidad Metropolitana, and the
Universities Space Research Association.

Facilities: \facility{Planck}, 
\facility{Arecibo},
\facility{IRAS}

\bibliographystyle{apj}
\bibliography{wtrbib}

\end{document}